\documentclass[aps,prl,twocolumn,showpacs,groupedaddress]{revtex4}  
\usepackage{graphicx}  
\usepackage{dcolumn}   
\usepackage{bm}        
\usepackage{amssymb,amsmath}   
\usepackage{psfrag}
\usepackage{epsfig}
\usepackage{dsfont}

\newcommand{\DD}{\mathcal D}
\newcommand{\dd}{\mathrm d}
\newcommand{\Det}{\mathrm{Det}}
\newcommand{\xUnit}{{\mathds 1}}

\newcommand{\C}{{\mathcal C}}

\newcommand{\wc}{{\omega_C}}

\newcommand{\epsth}{{\epsilon_{\mathrm{Th}}}}

\newcommand{\act}{{\mathcal A}}
\newcommand{\tm}{{\mathcal M}}

\newcommand{\vF}{{v_D}}

\newcommand{\sign}{\mathrm{sign}\,}

\newcommand{\nuEff}{\nu^\ast}

\begin{document}


\title{Influence of Coulomb interaction on the Aharonov-Bohm effect in an electronic Fabry-P\'erot interferometer}
\author{                                                                      
St\'ephane~Ngo~Dinh$^{1}$ and Dmitry~A.~Bagrets$^{2,3}$                                                             
}                                                                             
\affiliation{                                                                 
$^{1}$Institut f\"ur Theorie der \!Kondensierten \!Materie and DFG Center for Functional Nanostructures, Karlsruhe Institute of Technology, 76128 Karlsruhe, Germany\\
$^{2}$Institut f\"ur Nanotechnologie, Karlsruhe Institute of Technology, 76021 Karlsruhe, Germany\\
$^{3}$Institut f\"ur Theoretische Physik, Universit\"at zu K\"oln, Z\"ulpicher Str.~77, 50937 K\"oln, Germany
}                                                                             

\date{\today}

\begin{abstract}
We study the role of Coulomb interaction in an electronic Fabry-P\'erot interferometer (FPI) realized with
chiral edge states in the integer quantum Hall regime 
in the limit of weak backscattering.
Assuming that a compressible Coulomb island in a bulk region of the FPI is formed, we develop a 
capacitance model which explains the plethora of experimental data on the flux and gate periodicity
of conductance oscillations. It is also shown that a suppression of finite-bias visibility 
stems from a combination of weak Coulomb blockade and a nonequilibrium dephasing by the 
quantum shot noise.
\end{abstract}

\maketitle 

\begin{figure}[t]
	\centering{\includegraphics[width=2.2in]{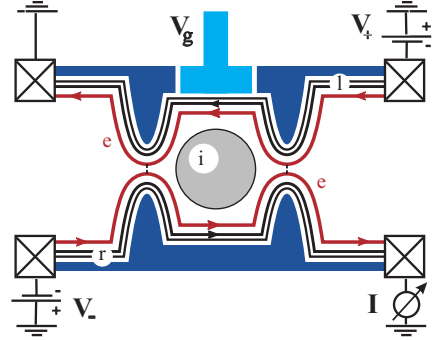}}
	\caption{Fabry-Perot interferometer with a center compressible island ``$\,i\,$''. 
The innermost edge channels ``$e$'' are subject to backscattering at the QPCs; the remaining $f_T=\nu-1$ right($r$)- and left($l$)- moving channels are fully transmitted. }
	\label{fig:Setup}
\end{figure}

Electron interferometry  provides a powerful tool for studying the quantum interference and dephasing in mesoscopic semiconductor devices. Notably, the 2DEG placed in the quantum Hall effect (QHE) regime 
has been proved to be highly tailored 
to realize electronic analogues of optical interferometers, such as 
Fabry-P\'erot (FPI)~\cite{MarcusWest09, ZhangMarcus09, Yamauchi09, Heiblum09} 
and Mach-Zehnder (MZI)~\cite{Heiblum03} interferometers,
with chiral edge states playing the role of light beams and quantum point contacts (QPCs) serving as beam splitters.  
These experimental efforts are motivated by the recent interest in topological quantum computations,
which propose to exploit the non-Abelian anyons in the fractional QHE regime~\cite{Nayak08}.   

\begin{table}[b]
\caption{\label{tab:table1}
``Phase diagram'' of the FPI (for definitions of 
inter-channel interaction $\alpha$ and capacitances $C_{gi}$, $C_{ei}$, see text). 
The AB regime in the limit of $\alpha\gg 1$ 
(``?'') was not observed.
}
\begin{ruledtabular}
\begin{tabular}{c|c|c}
 & ~~ $C_{gi}/C_{ei}\ll 1$~~ & ~~ $C_{gi}/C_{ei}\gg 1$ ~~ \\
\hline
&& \\
~ $\alpha \gg \ln{(w/a)}$, \, $\nu^*\simeq 1$  ~ & CD, type I &  ? \\
&& \\
~ $\alpha \ll \ln{(w/a)}$,\, $\nu^*\simeq \nu$ ~ & CD, type II & AB \\
&& \\
\end{tabular}
\end{ruledtabular}
\end{table}

The Coulomb interaction is of paramount importance in fractional QHE systems where it gives rise
to quasi-particles with fractional charge obeying anyonic statistics. It came as a surprise that {\it e-e}
interaction plays a prominent role in the integer QHE interferometers as well,
even when their conductance is $\sim e^2/h$
so that Coulomb blockade physics seems to be inessential. Experiments
realizing MZIs and FPIs in this limit can not be explained by means of the non-interacting Landauer-B\"uttiker 
scattering approach applied to chiral edge states.
The search for a resolution of this puzzle in the case of MZI has triggered a lot of attention. 
On the contrary, the extent of theoretical works on FPIs operating in the integer QHE regime is rather 
small~\cite{Chamon97,Halperin07, Halperin10},
and a theory for non-equilibrium edge-state dephasing in these devices has not been elaborated in detail.

In this paper we propose a minimal capacitance model of {\it e-e} interaction in the FPI and apply it
to study its transport properties in- and out-of-equilibrium in the limit of weak backscattering.
Our approach is inspired by the previous theoretical work~\cite{Halperin07}.
Its essential idea is that in the center of the FPI between two constrictions 
a compressible Coulomb island can be formed (Fig.~1) which
strongly affects Aharonov-Bohm oscillations.

 We show, that depending on the strength of
inter-channel {\it e-e} interaction $\alpha= e^2/(\hbar v_D)$, with $v_D$ being the drift velocity,
and on the ratio of gate-to-island ($C_{gi}$) to edge-to-island capacitance 
($C_{ei}$) the FPI can fall into  ``Aharonov-Bohm'' (AB) or ``Coulomb-dominated'' (CD)
regimes observed in the experiments~\cite{ZhangMarcus09,Heiblum09} (Table I),
including the ``exotic'' behavior discovered in~\cite{Heiblum09}, which we classify as the type II CD regime.  
There is a partial overlap 
of our results at equilibrium with recent work~\cite{Halperin10}.
We also analyze the suppression of  AB oscillations out-of-equilibrium with the increase of a source-drain 
voltage and find both power-law and exponential decays, which may explain experiments of 
Refs.~\cite{MarcusWest09, Yamauchi09}. 

\emph{Model} --- We consider an electronic FPI of size $L$ formed by a Hall bar 
with $\nu$ edge channels 
and two constrictions (QPCs) that allow for electron backscattering between the innermost right-/left-moving 
edge channels with amplitudes $r_{1(2)}$ as shown in Fig.~\ref{fig:Setup}. 

According to 
Chklovskii {\it et al.}~\cite{Chklovskii92}, the 
2DEG in the QHE regime is divided into compressible and incompressible strips 
(if $w$ and $a$ are their typical widths, 
then $w\gg a \gg \lambda_B$, with $\lambda_B$ being the magnetic length). 
Compressible regions play the role of edge channels separated
by narrow insulating stripes of the incompressible 2DEG. We assume that  
the filling fraction in the center of the FPI exceeds $\nu$ giving rise to  
a compressible droplet (Coulomb island). The reason for that can be smooth (on a scale $\lambda_B$) disorder 
potential fluctuations~\cite{Cooper93}.

We 
take {\it e-e} interaction in the FPI into account 
using the constant interaction model with mutual capacitances between four 
compressible regions ---
the interfering channel ($e$); 
right- and left-moving fully transmitted channels ($r$, $l$); the compressible island ($i$) ---
and the gate ($g$).
One estimates them as $C_{r(l)e}\sim\epsilon L \ln(w/a)$, $C_{r(l)i}\sim\epsilon r\ln(r/a)$,
$C_{eg} \sim\epsilon wL/d$ and $C_{ig}\sim \epsilon r^2/d$ 
with $d$ being a depth of 2DEG below a top gate,
and we take the droplet to be a disk of size $r \simeq L$.
We assume a large capacitance between counter-propagating innermost channels ---
thus they share the same
electrostatic potential $\varphi_e$ --- and also consider $f_T$ right- and left-moving channels as 
joint conductors with potentials $\varphi_r$ ($\varphi_l$). 
Defining a capacitance matrix $\tilde C$ with elements $\tilde C_{\alpha\alpha}=\sum_\gamma C_{\alpha\gamma} + C_{\alpha g}$, and $\tilde C_{\alpha\beta}=-C_{\alpha\beta}$ ($\alpha\neq \beta$), 
where Greek indices span the set $\{e,r,l,i\}$ and $q_\alpha=-C_{\alpha g}V_g$ is an offset charge on the $\alpha$'s conductor, 
the electrostatic energy reads
\begin{equation} 
\label{eq:E_int}
E = (1/2)\sum_{\alpha \beta} \left(Q_\alpha-q_\alpha\right) \left(\tilde C^{-1}\right)_{\alpha\beta} \left(Q_\beta-q_\beta\right).
\end{equation}
Here the total charge on the island, $Q_i=e\left(N_i+\nu \phi/\phi_0\right)$, is contributed by the highest partially filled Landau level (LL) and $\nu$ fully occupied underlying LLs~\cite{Halperin07}. 
With $N_i$ being integer, the former part of this charge is quantized; $\phi$ stands for a magnetic flux through the FPI and $\phi_0=hc/\lvert e\rvert$ is the flux quantum (we take a convention $e<0$).

\begin{figure}[t]
\includegraphics[scale=0.25]{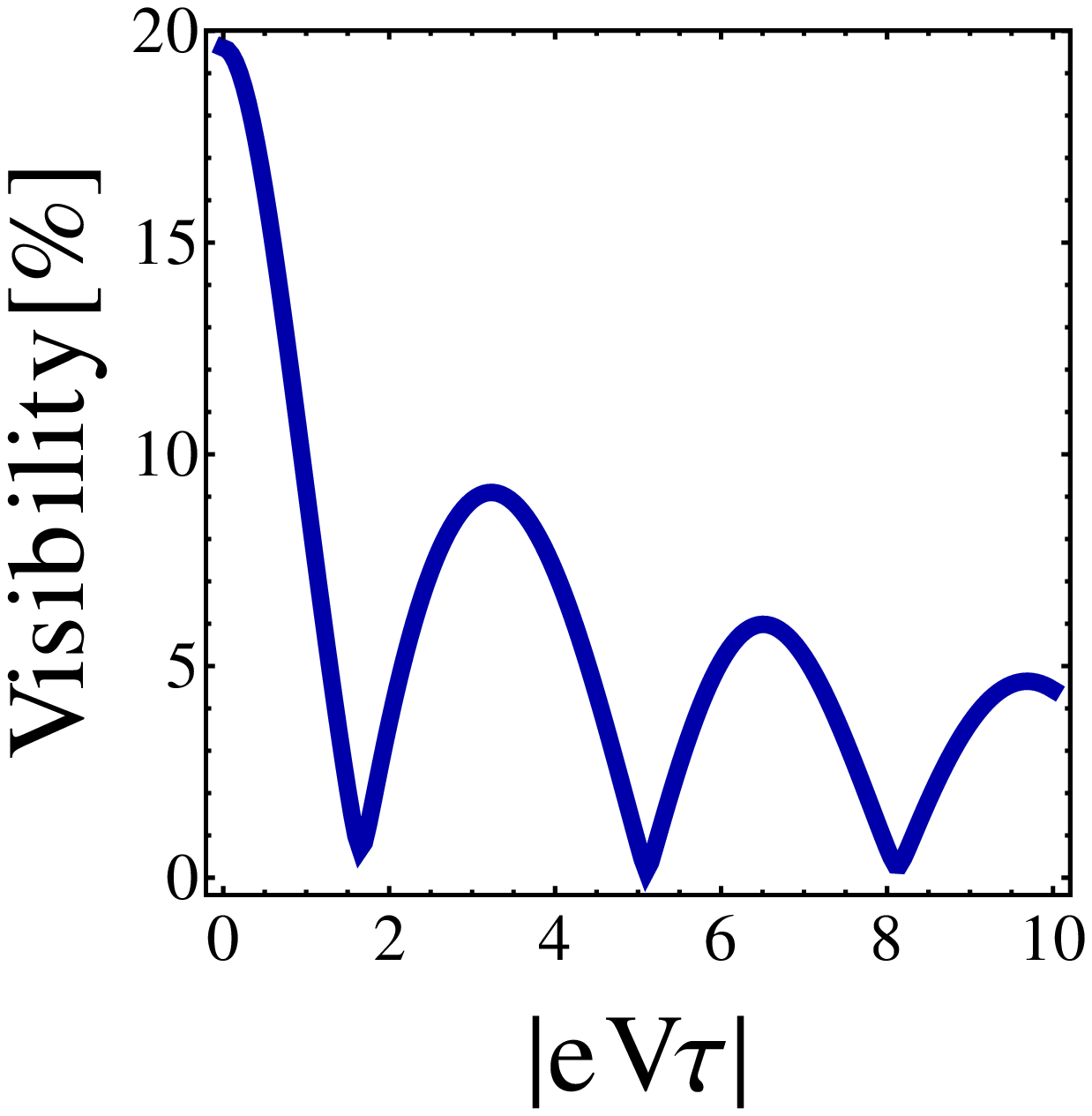}	
\includegraphics[scale=0.3]{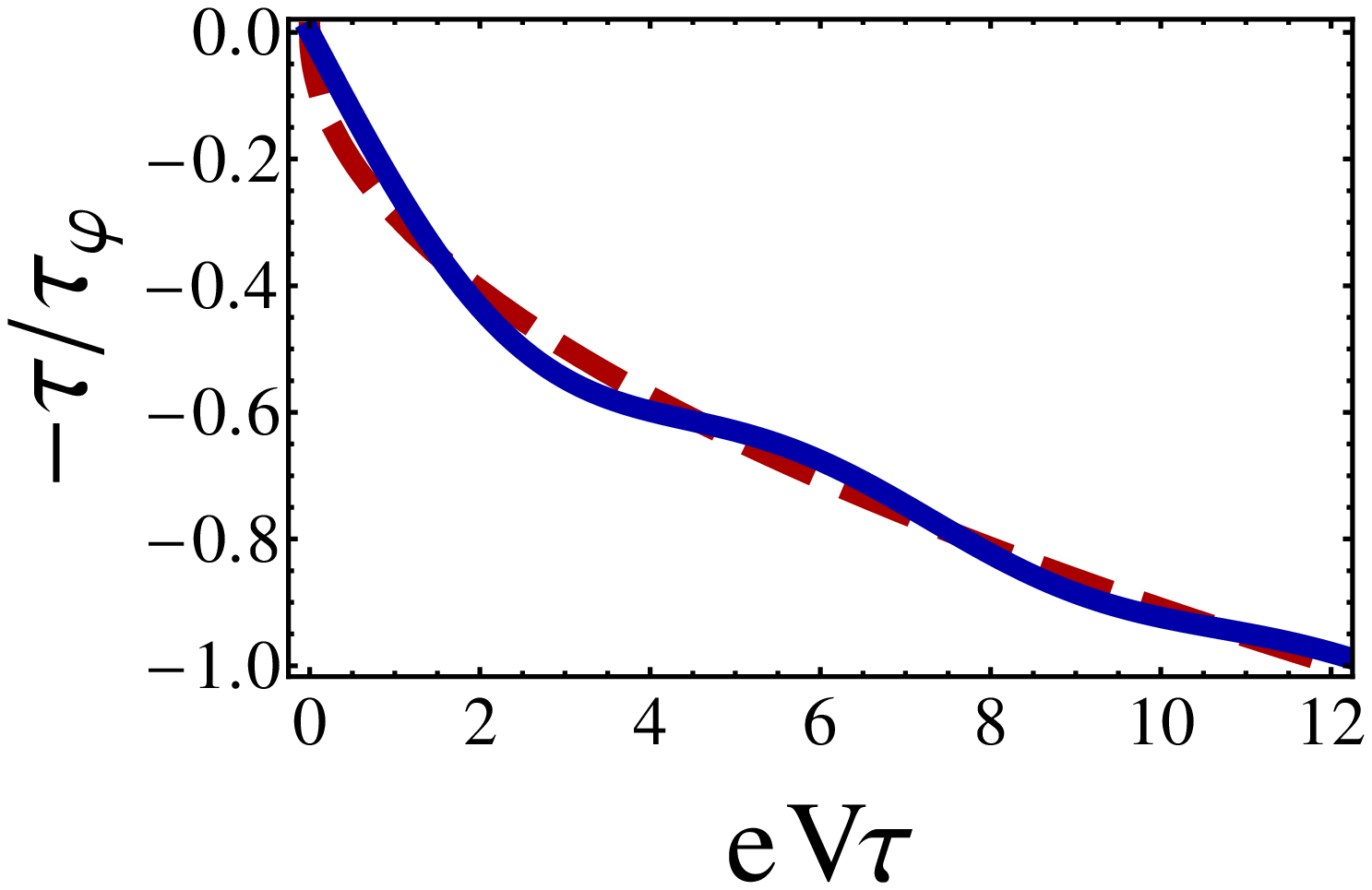}
\caption{(Left) visibility of AB oscillations, $g(V)/\nu$;
(Right) dephasing rate as a function of source-drain voltage,
the dashed line is the power-law asymptotic given by Eq.~(\ref{eqn:TauPhi}).
Parameters are $\nuEff=2$, $\wc\tau=25$, $R_{1\ast}(\epsth)=R_{2\ast}(\epsth)=0.2$.}
\label{fig:GGateBias}
\end{figure}

\emph{Results} - In the weak backscattering limit $r_j\ll 1$, the dependence of the AB conductance 
on external parameters --- the gate voltage $V_g$, the variation of magnetic field $\delta B$ and 
the bias $V=V_+ - V_-$ ---
factorizes into 
\begin{equation}
\label{eqn:gAB}
g_{AB}(V,V_g,\delta B) = g(V) \cos\left[\varphi_{AB}(V_g,\delta B)\right],
\end{equation}
where the AB phase $\varphi_{AB}$ will be discussed shortly and 
\begin{equation} \label{eqn:gV}
 g(V) = 2 e^{-\tau/\tau_\varphi(V)} R_{12*}(V)\, \left|\cos\left(|eV \tau| - {\pi}/{4\nu^*}\right)\right|.
\end{equation}
Here $\tau=L/v_D$ is the flight time of electrons through the FPI, 
 $\nuEff$ is the effective number of channels participating in screening -- 
$\nuEff\simeq 1$ in the case of strong inter-channel coupling $\alpha\gg \ln(w/a)$ while 
$\nuEff\simeq \nu$ in the weak coupling limit --, and the
nonequilibrium dephasing rate   
\begin{equation} \label{eqn:TauPhi}
	\tau_\varphi^{-1} = (2/\pi) \lvert eV\rvert \bigl(R_{1\ast}(V)+R_{2\ast}(V)\bigr)\sin^2(\pi/{2\nuEff})\, .
\end{equation}
There are two characteristic energy scales in our problem, $\epsilon_{\rm Th} = \pi/\tau$ being 
the Thouless energy and $\omega_C$ being the charge relaxation rate, 
where  $\epsilon_{\rm Th} \ll \omega_C$.
In the range $\epsilon_{\rm Th} \ll |eV| \ll \omega_C$ the
renormalized reflection coefficients 
behave as $R_{j\ast} \propto V^{-1/\nuEff}$ and $R_{12\ast} \propto V^{-1/2\nuEff}$
where one sets $R_{j\ast} \simeq |r_j|^2$ and $R_{12*} \simeq |r_1 r_2|$ 
at high energy scale $\omega_C$.
This renormalization comes from {\it virtual} electron-hole excitations  
(being a precursor of weak Coulomb blockade~\cite{Matveev95}) and stops at $eV \simeq \epsth$, 
whereas $\tau_{\varphi}^{-1}$ is caused by {\it real} e-h pairs excited by
backscattered electrons and is proportional to the shot noise of the QPCs.
As the function of bias, the amplitude~(\ref{eqn:gV}) 
of AB oscillations shows the ``lobe'' structure (Fig.~\ref{fig:GGateBias}) 
 on a scale of the Thouless energy $\epsth$
in agreement with Refs.~\cite{MarcusWest09, Yamauchi09}.

\begin{figure*}[t]
\centering{
\includegraphics[scale=0.22]{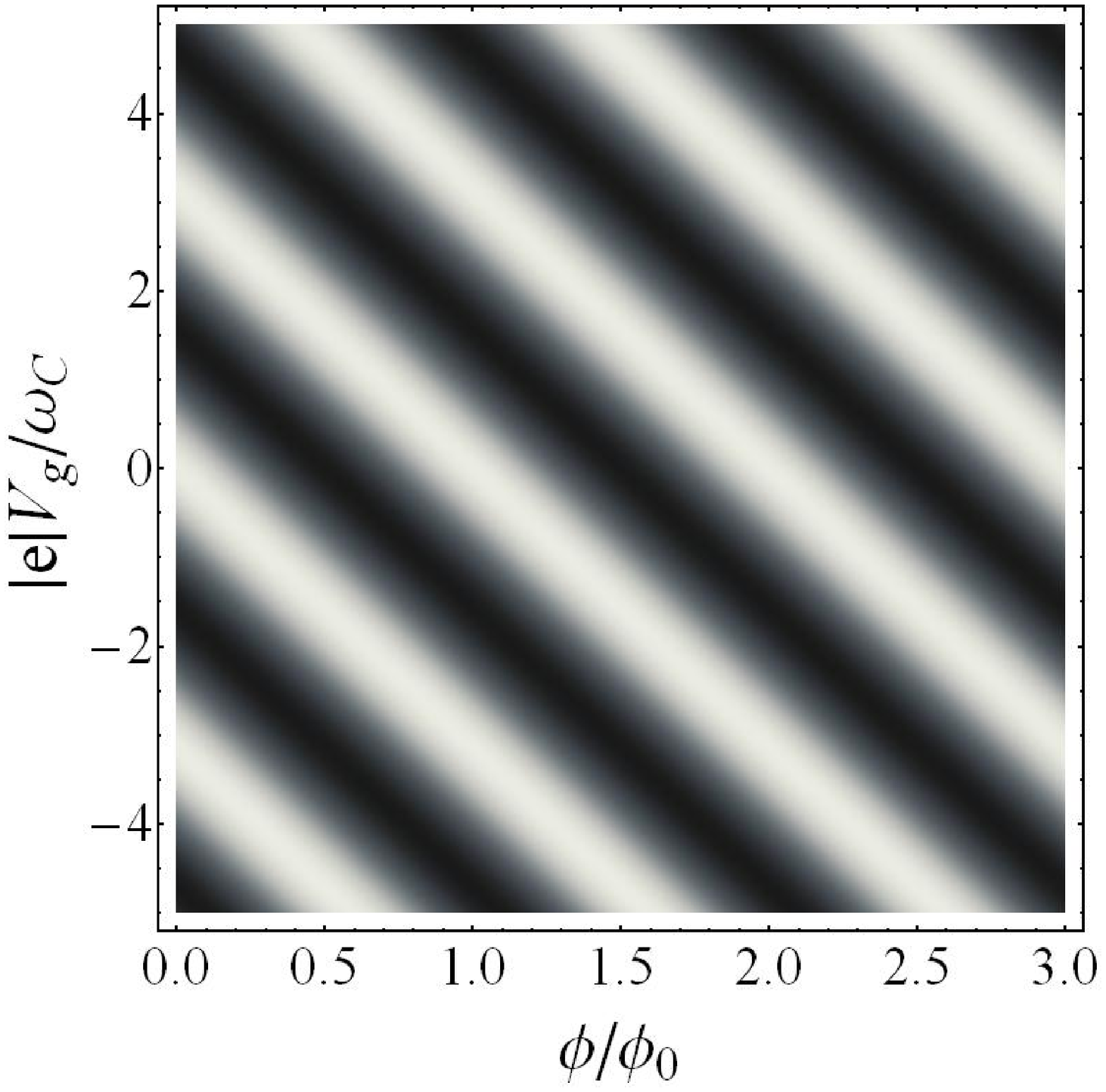}
\includegraphics[scale=0.22]{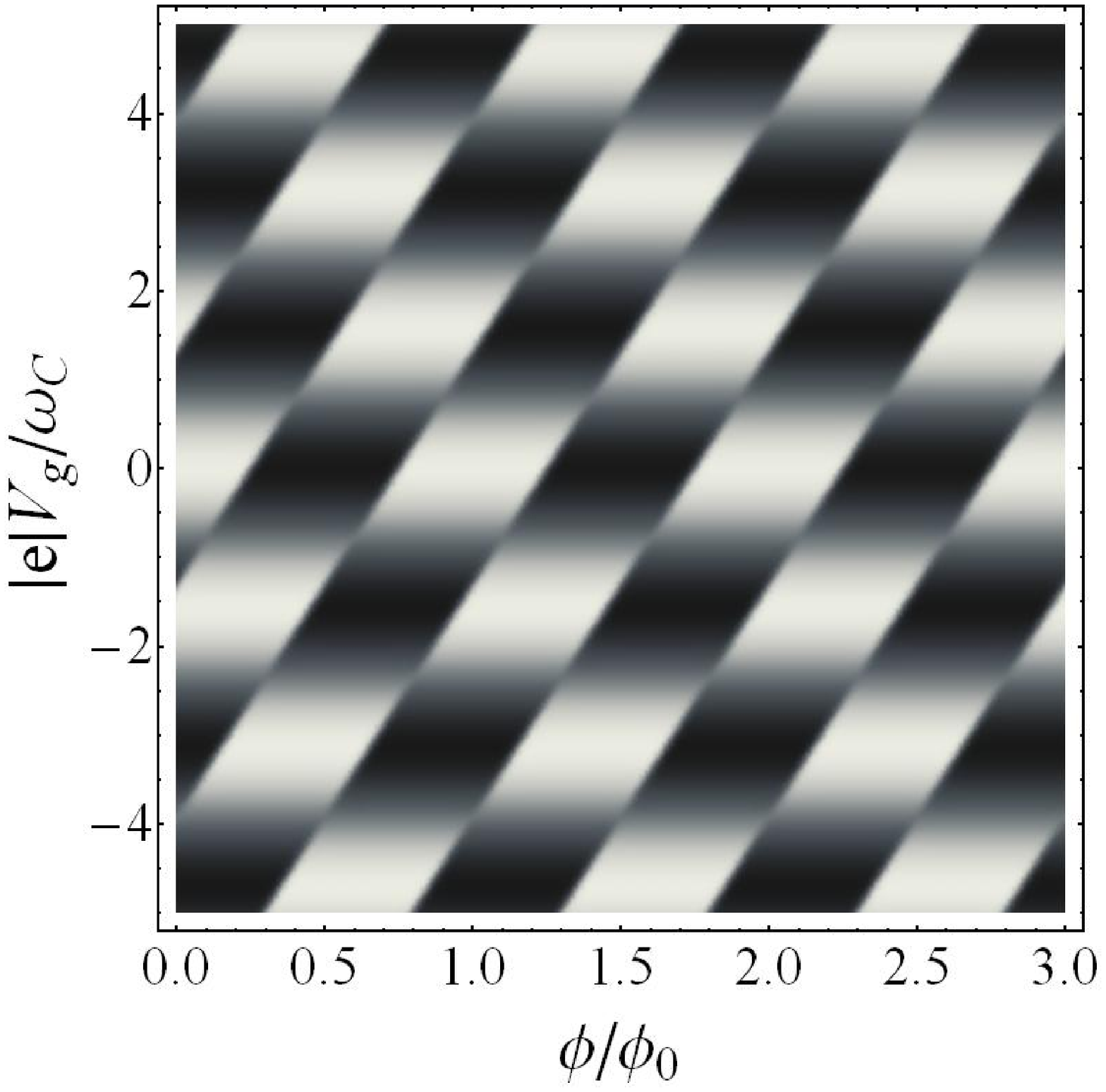}
\includegraphics[scale=0.22]{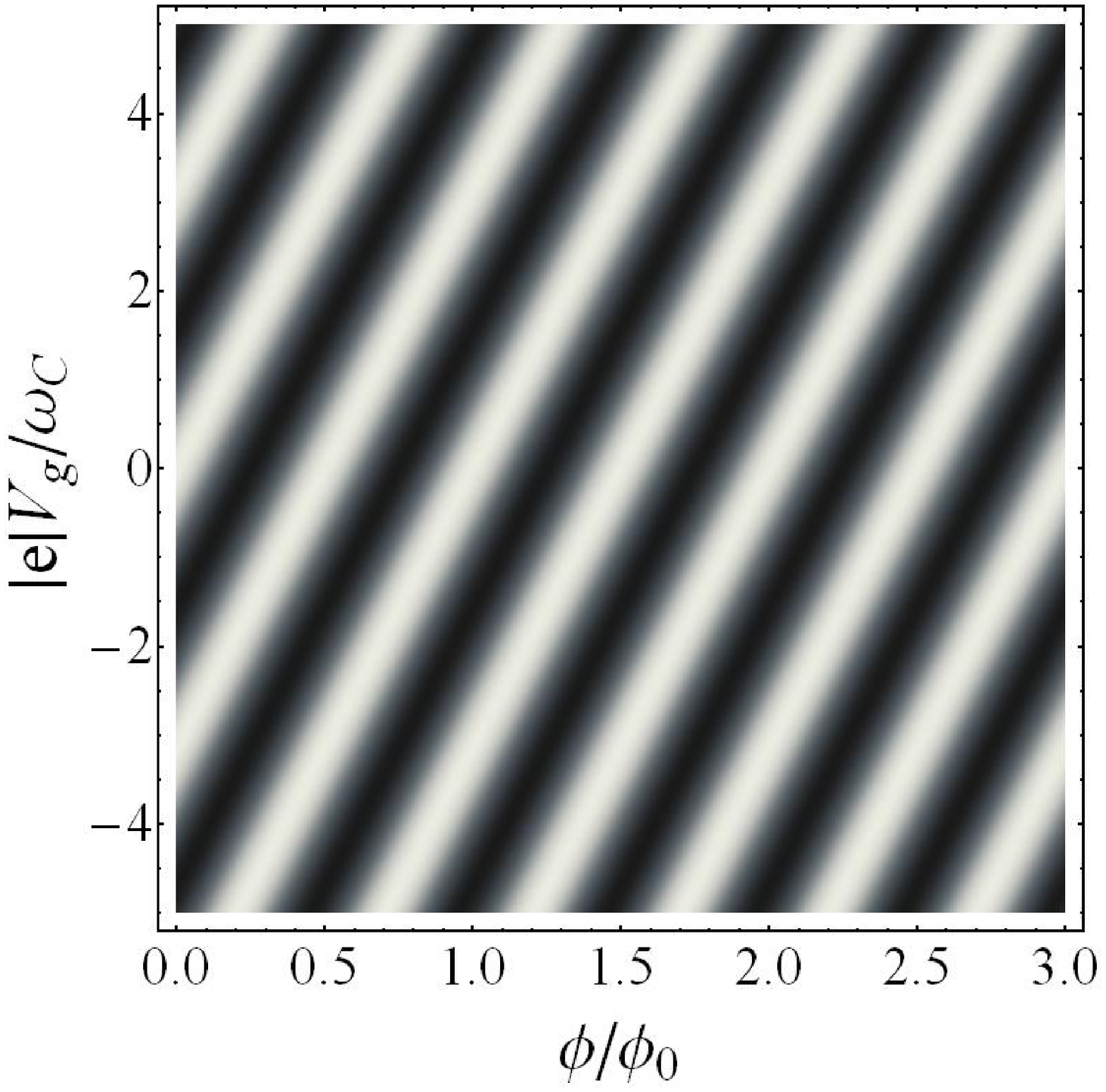}
}
\caption{AB conductance: 
AB regime (left), $\nu=2$; type-II CD  (middle), $\nu=2$, $C_{ig}/\bar C_e = 0.6$; type-I CD (right), 
$\nu=3$.  }
\label{fig:GNoIStrongEE}
\end{figure*}

In experiment one usually characterizes the FPI in terms of a pattern of its
equilibrium conductance in $(B,V_g)$-plane, which is governed by AB phase. 
Defining island and edge capacitances as $\bar C_i=\bar C_{ei} + C_{ig}$ and 
$\bar C_e = \bar C_{eg} + \bar C_{ei} C_{ig}/\bar C_i$ respectively, where 
$\bar C_{ei} = C_{ei} + C_{ri} + C_{li}$ (and similarly for $\bar C_{eg}$), 
we have found that
\begin{eqnarray}
\varphi_{AB} &=& 2\pi \phi/\phi_0 - \frac{2\pi}{\nu^*}\frac{\bar C_{ei}}{\bar C_i}\left(N_i + \nu \phi/\phi_0\right)
\nonumber \\
&+& |e|\left(2V_g-V_+-V_-\right)/\omega_C.
\label{eq:ABphase}
\end{eqnarray}  
Here $\phi=A\delta B$ is the variation of the flux through some reference loop area,
$\omega_C = \nu^* e^2/(\pi\,\bar C_e)$ is the charge relaxation rate, and $N_i$ minimizes
the charging energy
\begin{equation}
\label{eq:Ei}
E_i = \frac{e^2}{2\bar C_i}\Bigl(N_i + \nu \phi/\phi_0 - C_{ig}V_g/|e|\Bigr)^2.
\end{equation}

\emph{Discussion} - Importantly, Eq.~(\ref{eq:ABphase}) can be interpreted in terms of variation of the relevant FPI's area 
when the magnetic field and voltage are varied~\cite{Heiblum09}. Representing the magnetic flux 
as $\Phi = B \cdot A(V_g, B)$, this leads to
\begin{equation}
 \left(\frac{\partial A}{\partial B}\right)_{V_g}\! = - \frac{\nu}{\nu^*}
\left(\frac{\bar C_{ei}}{\bar C_i}\right)\frac{A}{B},\quad
 \left(\frac{\partial A}{\partial V_g}\right)_{B} = \frac{|e|}{\omega_C}\frac{\phi_0}{\pi B},
\end{equation}
i.e.\ a qualitative dependence of $A(B,V_g)$ on the parameters $\nuEff$ and $\bar C_{ei}/\bar C_i$ and hence on the different regimes of Table I. 
The corresponding patterns of AB conductance are shown in Fig.~\ref{fig:GNoIStrongEE}. 

In the AB regime 
the magnetic field period is $\Delta B = \phi_0/A$ and
the area of the FPI does not change with $B$ yielding the lines of constant phase with
a negative slope (Fig.~\ref{fig:GNoIStrongEE}, left). 
This regime is observed in large devices (cell area $\sim$20$\mu{\rm m}^2$) 
with a top gate~\cite{MarcusWest09, ZhangMarcus09, Heiblum09}, where the condition 
$\bar C_{ei} \ll C_{gi}$ is satisfied. Assuming that $\bar C_{e} \simeq \bar C_{ei} + \bar C_{eg}$ 
is $\nu$-independent, one obtains a gate period $\Delta V_g = \pi\omega_C/e \propto 1/B$, 
consistent with Ref.~\cite{ZhangMarcus09}. 

To distinguish between type-I and II CD regimes we compare an inter-channel interaction energy
$e^2/C_{re}$ with the Thouless energy $\sim v_D/L$.
Using the estimate for $C_{re}$ one gets a crossover value for the coupling constant
$\alpha^* \sim \ln (w/a)$. In the CD regime one has $\bar C_{ei}/\bar C_{i}\simeq 1$ and the area of the interfering loop 
shrinks with the increase of magnetic field. In the type-II regime such shrinkage exactly compensates
a change in ``na\"ive'' AB phase $\phi=A\delta B$, which causes the true AB phase to stay piecewise constant
while keeping $V_g$ fixed. When the FPI is brought close to a charge degeneracy point of the island
by varying $V_g$ or $B$, electron tunneling becomes possible between the droplet and
interfering channels resulting in abrupt change of $A$. This creates a phase lapse
$\Delta\phi_{AB} = \pm 2\pi/\nu$ giving rise to the ``rhomb-like'' pattern shown
in Fig.~\ref{fig:GNoIStrongEE} (middle) 
 at $\nu \ge 2$. 
Such ``exotic'' behaviour of  AB oscillations has been reported in Ref.~\cite{Heiblum09}.

In the type-I CD regime a change in AB phase caused by area shrinkage when rising $B$
overcompensates the ``na\"ive'' AB phase. At the same time, whenever an electron tunnels into the island,
the interfering edges contract so as to expell exactly one flux quantum from the AB loop
(the phase lapse $\Delta\phi_{AB} = \pm 2\pi$), which is invisible in the interference conductance. 
This leads to the diagonal stripe pattern  (Fig.~\ref{fig:GNoIStrongEE}, right) 
 with
periodicities $\Delta B = \phi_0/(f_T A)$ and $\Delta V_g = e/(C_{ei} + C_{gi})$
and lines of constant phase having positive slope (at $\nu=1$ lines are vertical)  (cf. Ref.\cite{Halperin07}) 
The above scenario has been realized in
Refs.~\cite{ZhangMarcus09, Heiblum09} for small FPIs (area $2 \div 5 \mu{\rm m}^2$),
where a presence/absence of a top gate did not affect their behavior,
suggesting the CD limit $C_{ei} \gg C_{gi}$.

\emph{Calculations} - We now turn to sketch our calculations the details of which will be published separately.
The FPI is modeled by 1D chiral fermions, interacting according to Eq.~(\ref{eq:E_int}). We use the
functional bosonization framework~\cite{Grishin04} and decouple interaction 
by means of electrostatic potentials $\varphi_{\alpha}(t)$
with Lagrangian 
\begin{eqnarray}
{\cal L}_{0}\!=\!\sum_{\langle \alpha\beta\rangle}\frac{C_{\alpha\beta}}{2}(\varphi_\alpha-\varphi_\beta)^2
\!+\!\sum_{\alpha\neq g} \frac{C_{\alpha g}}{2}(\varphi_\alpha-V_g)^2\!-\!\varphi_i Q_i
\end{eqnarray}
Along the lines of Ref.~\cite{SnymanNazarov} one can now integrate out the fermions
to obtain the Keldysh action describing electron scattering in the FPI:
\begin{equation}\label{eqn:levitov}
	i\act_{\rm FPI}[\varphi] = \ln \Det \left[\xUnit+\left(S[\varphi^b]^\dagger S[\varphi^f]-\xUnit\right)f\right].
\end{equation}
This result bears close relation to the problem of electron full counting statistics~\cite{Levitov96}. 
The determinant is to be taken with respect to time and channel indices $\mu$; 
$f_{\lambda\mu}=\delta_{\lambda\mu} f^<_\mu = ie^{-ieV_\mu t}/2\pi(t+ i 0)$ are zero temperature 
occupation numbers in the incoming channels 
and $S_{\lambda\mu}[\varphi^{f/b}](t,t')$ is the time-dependent single-particle
scattering matrix of the FPI in the potential $\varphi^{f/b}$ on two branches of the Keldysh contour $\mathcal C$. 
It can be constructed using the scattering matrices of the QPCs and the transfer matrix
$\tm_\pm(t,t_0) = e^{i\theta_\pm(t)+ i\phi_\pm} \delta(t-t_0-\tau)$ of electron
along the arm of the FPI between points of scattering $x_{1,2}$ (see Appendix).
Accumulated phases due to electric and magnetic fields satisfy $2\pi\phi/\phi_0=\phi_++\phi_-$
and
\begin{equation}
\label{eq:FBphase}
\theta^{f(b)}_\pm(t)=\mp v_D^{-1} \int_{x_1}^{x_2}\!\dd x'\, e\varphi^{f(b)}(t-(x_{2/1}-x')/v_D).
\end{equation}

We have analyzed the problem in the weak backscattering limit and $T=0$ performing the expansion 
of $\act_{\rm FPI}[\varphi]$ up to order of $|r_j|^2$ in a similar fashion as in Ref.~\cite{Schneider11}. 
The intermediate result is quadratic in variables $\varphi_{r(l)}$ and $\varphi_i$, 
and they can be integrated out. The resulting action then takes
a form $\act_e[\varphi_e]= \act_{RPA} + \act_b$, where
\begin{multline} 
\label{eqn:SInt}
	\act_{RPA} = \frac 12 \int_\C\!dt_{1,2}\,\varphi(t_1) V^{-1}(t_1-t_2) \varphi(t_2) \\
    - \int_\C\!\dd t\,\varphi(t) Q_0(t)-\frac 12 \bar C_i^{-1} \int_\C\!\dd t\, \left(Q_i-q_i\right)^2
\end{multline}
with
\begin{equation}
\label{eq:V_RPA}
	V^{r/a}(\omega) = \frac{\omega/\bar C_e}{\omega\pm i \wc\left(1-e^{\pm i\omega\tau}\right)}
\end{equation}
being the effective  RPA {\it e-e} interaction and the charge 
\begin{equation}
\label{eq:Q0}
Q_0 = \frac{\nu^* e^2 \tau}{2\pi}(V_+ + V_-) + \bar C_e V_g + \frac{\bar C_{ei}}{\bar C_i} Q_i.
\end{equation}
The last term of Eq.~(\ref{eqn:SInt}) is just the electrostatic energy~(\ref{eq:Ei}) of the Coulomb island. 
The backscattering action (being non-linear in $\varphi_e$) reads
\begin{equation} \label{eqn:Sb}
	i\act_b = \sum_{kl} \int_\C dt_{1,2} e^{-2i \Phi(x_k,t_1)} \Pi_{kl}(t_1-t_2) e^{2i \Phi(x_l,t_2)}
\end{equation}
with the polarization operators defined as
\begin{equation} \label{eqn:backpol}
\Pi^\gtrless_{kl}(t) = - r_k r_l^* e^{- 2i\pi\epsilon_{kl}\phi/\phi_0}
 f_+^\gtrless(\epsilon_{kl}\tau+t) f_-^\lessgtr(\epsilon_{kl}\tau-t), 
\end{equation}
where $\epsilon_{kl}$ is the antisymmetric tensor, $f_\mu^>(t)=\delta(t)-f^<_\mu(t)$,
and the relative phase 
$\vec{\Phi}~=~\hat{\mathcal D}\sigma_3\vec{\varphi_e}$.
Here we have introduced doublets, e.g. 
$\vec{\varphi_e} = (\varphi_e^f, \varphi_e^b)^T$, and the particle-hole propagator has the conventional
structure in Keldysh space with
$\mathcal D^{r/a}(\omega;x,x')=\sign (x-x')\ e^{\pm i \omega\lvert x-x'\rvert/\vF}/2\vF$.

\begin{figure}[t]
 	\centering{\includegraphics[scale=0.3]{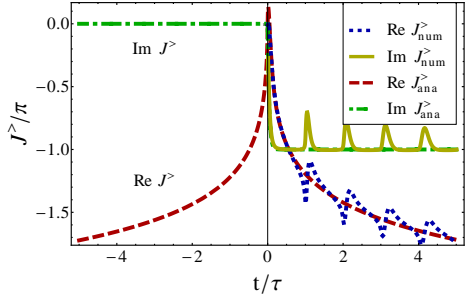}}
 	\caption{Correlation function $J^>$: comparison of analytic approximation $J^>_{\rm ana}$ and numerical results $J^>_{\rm num}$ for $\wc\tau=25$.}\label{fig:JNumAna}
 \end{figure}
The backscattering current can be now obtained using the relation 
$I_b=e\sum_{ij} (I_{ij}^>-I_{ij}^<)$, where the forward/backward backscattering rates 
are represented as the path integral over $\varphi_e$,
\begin{equation} 
I^{\gtrless}_{ij}\!\!=\!\!\int\! \dd t\,\DD\varphi_e\, e^{i\act_e[\varphi_e]-2 i \Phi^{b/f}(x_i,t) }
\Pi^{\alpha\beta}_{ij}(t) e^{ 2i\Phi^{f/b}(x_j,0) }. 
\label{eqn:Ib}
\end{equation}
The evaluation of this integral requires knowledge of the phase correlator 
$D_\Phi\equiv-i\left\langle \Phi \Phi \right\rangle=-\mathcal DV\mathcal D$. Using the RPA 
interaction~(\ref{eq:V_RPA}) and the definition of the p-h propagator, given above, one readily obtains
\begin{equation}
D^\gtrless_\Phi(t;x_i,x_j) =  \frac{ i A_{ij}}{8\nu^*}\{ J^\gtrless(t-\tau)\!-\!J^\gtrless(t)+
(t\rightarrow -t)^*\}, 
\end{equation}
where we have denoted $A_{ij}=2\delta_{ij}-1$ and
\begin{equation} \label{eqn:JIntegral}
  		J^\gtrless(t) \equiv  \int_0^{\pm\infty} \!\! \frac{d\omega}{\omega} \frac{i\wc \left(1-e^{i\omega \tau}\right)}{\omega+i\wc \left(1-e^{i\omega \tau}\right)} \left(e^{-i\omega t}-1\right).
\end{equation}
This integral is accessible only numerically  (Fig.~\ref{fig:JNumAna}). 
However, in the long time limit, $\wc t \gg 1$, 
which is consistent with a low bias condition $eV \ll \omega_C$,
one can use the approximation $J^\gtrless(t) \approx - \ln\left[-\wc (t\mp i a)\right]$ with 
a short time cutoff $a\sim \wc^{-1}$. 
Logarithmic correlations make this problem essentially similar to the one of 
tunneling into the non-equilibrium Luttinger liquid with impurity~\cite{Bagrets10}.
We have used the real-time instanton method \cite{Appendix11} developed in Ref.~\cite{Bagrets10} to evaluate the functional 
integral~(\ref{eqn:Ib}). It gives 
the AB phase~(\ref{eq:ABphase}) and the conductance~(\ref{eqn:gV}) 
with renormalized reflection coefficients and the 
bias dependent dephasing rate $\tau_\varphi^{-1}$ shown in Fig.~\ref{fig:GGateBias}(right).

\emph{Conclusions} --- We have presented the theory of electronic Fabry-Perot interferometer (FPI),
taking into account {\it e-e} interaction using the simple capacitive network model.
We have unraveled the mystery of the diversity in experimental observations of the AB effect 
in the FPI by 
presenting the two-parameter ``phase diagram'' of the device (Table~I) discriminating between 
``Aharonov-Bohm'' and ``Coulomb-dominated'' regimes. The  same model predicts 
the nonequilibrium dephasing rate and the lobe 
structure for the visibility in agreement with experiment.  

We thank I.V.~Gornyi, M.~Heiblum, A.~Mirlin, N.~Ofek, D.G.~Polyakov, S.~Carr, and A.~Stern for
useful discussions.  This work was supported by EUROHORCS/ESF, by GIF Grant No.\ 965 and by CFN/DFG.

\section{Supporting Online Material}
\subsubsection{Scattering matrix of the FPI}
In order to construct the scattering matrix in Eq.~(\ref{eqn:levitov}) we need to identify the two main constituents of single-particle dynamics:
\begin{itemize}

\item[(a)] scattering at QPC $j$, $s_j=\begin{pmatrix}
                                     	t_{j++} & r_{j+-}\\
					r_{j-+} & t_{j--}
                                    \end{pmatrix}
$, $j=1,2$;
\item[(b)] propagation in between, along the lower/upper edge,
$\tm_\pm(t,t_0) = e^{i\theta_\pm(t)+ i\phi_\pm} \delta(t-t_0-\tau).$
\end{itemize}
Summing up all possible trajectories then gives the total scattering matrix $S[\varphi^\alpha]=\begin{pmatrix}
                                     	T_{++} & R_{+-}\\
					R_{-+} & T_{--}
                                    \end{pmatrix}$.
Electrons can encircle the interferometer an arbitrary number of times, each turn giving rise to an amplitude of $r_{1+-}\tm_-r_{2-+}\tm_+$. The corresponding geometric series gives $R_\infty=\left[1-r_{1+-}\tm_-r_{2-+}\tm_+\right]^{-1}$ such that e.~g.
\begin{align*}
	R_{-+} &= r_{1-+}+t_{1--}\,\tm_-\,r_{2-+}\,\tm_+\, R_\infty\, t_{1++},\\
	T_{++} &=t_{2++} \tm_+ R_\infty t_{1++},
\end{align*}
where the dependence of $\tm_\pm$ and hence of $S$ on the configuration $\varphi$ has been left implicit.


\subsubsection{Saddle point approximation}
We use the real-time instanton method developed in Ref.~\cite{Bagrets10} to evaluate the functional integral (\ref{eqn:Ib}). Optimizing the tunneling action 
\begin{equation}
\act_{ij} = \act_e[\varphi_e]-2i\Phi^\alpha(x_i,t_1,[\varphi_e])
+2i\Phi^\beta(x_j,t_2,[\varphi_e])
\end{equation}
with respect to $\varphi_e$ one obtains the approximate saddle-point $\varphi^\ast_e$, 
and the related phase
\begin{multline} \label{eqn:SP}
	\Phi^\gamma_\ast(t,x) = \left<\Phi^\gamma(t,x)\left[-2 i \Phi^\alpha(t_1,x_i)+2 i \Phi^\beta(t_2,x_j)\right]\right>\\+\Phi_0(x)
\end{multline}
where the dependence on $\alpha$, $\beta$, $x_i$, $x_j$, $t_1$, and $t_2$ has been left implicit. The phases $\Phi_0(x_{1,2})$ are accumulated by electrons in the mean electrostatic potential on the 
interfering channels,
\begin{equation} 
\Phi_0(x_{1,2}) = \mp\frac{\pi Q_0}{\nuEff e}\left(\frac{\wc\tau}{1+\wc \tau}\right),
\label{eqn:MFP}
\end{equation}
where the charge $Q_0$ was defined in Eq.~(\ref{eq:Q0}).

A substitution of the stationary phase (\ref{eqn:SP}) into the tunneling action $\act_{ij}$ 
renormalizes the backscattering polarization operators in Eq.~(\ref{eqn:Ib}), 
\begin{equation}
\tilde \Pi^{\gamma\delta}_{kl}	(t)  = \Pi_{kl}^{\gamma\delta}(t)
\exp\left\{-2\left<\left(\Phi^\gamma_k(t)-\Phi^\delta_l(0)\right)^2\right> \right\},
\end{equation}
thereby changing their singular behaviour to
\begin{equation}
\tilde \Pi_{kl}(t)\sim e^{-ieVt} t^{-2-2\lambda} (t^2-\tau^2)^\lambda.
\end{equation}
The exponents here are $\lambda=-1/2\nuEff$ for $k=l$ and $\lambda=1/2\nuEff-1$ for $k\neq l$ 
(thus, they assume their noninteracting values in the limit $\nuEff\to\infty$). 
To obtain the same re\-nor\-ma\-li\-za\-tion in the backscattering action (\ref{eqn:Sb}) 
one should take into account the quantum fluctuations around the saddle-point trajectory $\Phi_*^\gamma$. 

We concentrate on the limit of large bias, $\wc\tau\gg\lvert eV\tau\rvert \gg 1$, where the singularities of
the ``dressed'' polarization operator $\tilde\Pi(t)$ dominate the real-time integrals in 
Eqs.~(\ref{eqn:Sb}), (\ref{eqn:Ib}) and restrict ourselves to the case $\nuEff>1$. For each of the 
interference current contribution, $I_{12}$ and $I_{21}$, the most essential part of the integral
come from two singularities at $t \simeq \pm\tau$. Summing up these four terms one obtains 
the result~(\ref{eqn:gAB}) with the AB phase
\begin{equation}  
\varphi_{AB} = 2\pi\phi/\phi_0 + e(V_+ + V_-)\tau + \Phi_0(x_1)-\Phi_0(x_2).
\end{equation}
Using the relation~(\ref{eqn:MFP}) one can verify that in the limit $\omega_C\tau \gg 1$
which we consider the phase $\varphi_{AB}$ indeed coincides with the one given by Eq.~(\ref{eq:ABphase}).
 
On the other hand, the backscattering action evaluated on the saddle-point $\Phi_*^\gamma(t,x)$
is dominated by the incoherent terms $k=l$ around $t_1\simeq t_2$ in Eq.~(\ref{eqn:Sb}). 
They give rise to 
out-of-equilibrium dephasing due to quantum shot noise, $i\act_b[\varphi^\ast_e] =-\tau/\tau_\varphi$, 
with the voltage dependent rate $\tau_\varphi^{-1}$ shown in Fig.~\ref{fig:GGateBias}(right).

Incoherent corrections to the current, $I_{ii}$, are dominated by the singularities of Eq.~(\ref{eqn:Ib})
at $t\to 0$, yielding the total classical conductance
\begin{equation} 
g_{\rm cl} = \nu-\left(R_{1\ast}(V)+R_{2\ast}(V)\right).
\end{equation}
The applicability of the weak backscattering limit requires the renormalized 
reflection coefficient to satisfy  $R_{j*}(\epsilon_{\rm Th}) \ll 1$, which means
that the Coulomb blockade on the FPI has not yet developed. This gives a condition
$|r_j| \ll 1/(\omega_C \tau)^{1/2\nu^*}$ on initial values of the backscattering amplitudes 
in our model. 
\end{document}